# Hydrogen Bond Interaction Promotes Flash Energy Transport at MXene-water Interface

Jiebo Li[†,#], Zhen Chi[‡,#], Ruzhan Qin[§], Li Yan[∥], Xubo Lin[†], Mingjun Hu[∥,*], Guangcun Shan[§,*], Hailong Chen[‡,*], Yu-Xiang Weng[‡]

[†]Beijing Advanced Innovation Center for Biomedical Engineering, School of Biological Science and Medical Engineering, Beihang University, Beijing 100191, China

[‡]Beijing National Laboratory for Condensed Matter Physics, CAS Key Laboratory of Soft Matter Physics, Institute of Physics, Chinese Academy of Sciences, Beijing, 100190, China

[§] School of Instrument Science and Opto-electronics Engineering, Beihang University, Beijing, 100191, China;

[∥]School of Materials Science and Engineering, Beihang University, Beijing, 100191, China;

[⊥] Songshan Lake Materials Laboratory, Dongguan, Guangdong 523808, China

**\* Corresponding authors;**

Email: hlchen@iphy.ac.cn;  gcshan@buaa.edu.cn;  mingjunhu@buaa.edu.cn.

**Abstract**

There are emerging applications for photothermal conversion utilizing MXene, but the mechanism under these applications related interfacial energy migration from MXene to the attached surface layer is still unknown. In this paper, with comprehensive ultrafast studies, we reported the energy migration pathway from MXene ($Ti_3C_2T_x$) to local environment under plasmonic excitation. Our data found that in water, energy dissipation is divided into fast hydrogen bond mediated channel and slow lattice motion mediated channel. The experimental results suggest that in water, nearly 80% energy in MXene that gained from the photoexcitation quickly dissipates into surrounding water molecules within 7 ps as a hydrogen bond mediated fast channel, and the remaining energy vanishes with time constant ~100 ps as a lattice motion mediated slow channel. The fast energy migration would result in the prominent interfacial energy conductance 150-300 $MW \cdot m^{-2} \cdot K^{-1}$ for MXene-water interface. Tuning the solvent into ethanol could both narrow the energy dissipation to 35% through the fast channel and slow down the thermal channel (400 ps). To gain the molecular insight, molecular dynamic results presented different solvents had significantly different H bond forming ability on MXene surface. Our results suggested that interfacial interaction is crucial for effective hydrogen bonds on MXene surface to channel the excitation dissipation, providing important insights into the photothermal applications with MXene.

## 1. Introduction

MXene is a new structural unites two-dimensional material with promising potentials in many fields,[1, 2] such as energy storage and conversion,[3],[4] water treatment,[5] biomedical application,[6] catalysis,[7] electromagnetic interference shielding[8] and sensors.[9] In these versatile applications, the high electronic conductivity and plasmonic absorptions in near infrared ranges are the crucial to their excellent performances. In practical applications, the excellent intrinsic conductivity of MXene always associates with their reactive and hydrophilic surfaces, indicating that the interfacial interactions and energy migration on surface are also very important. For instance, the very recent paper reported that the cellulose interacted with MXene hydrophilic surface thus composed as flexible and conductive MXene films under near infrared light driven for smart actuator[10]. The fast responding speed of MXene-cellulose system under light stimuli is the critical step for this promising application. But, the physical process of thermal energy migration from MXene to cellulose inside this actuator is yet to be investigated. Similarly, due to its high photothermal conversion efficiency, MXene could be potentially used for water streaming,[11] cancer therapy[12-15] and biomedical imaging.[12, 16] However, there is still relatively poor understanding about the process how MXene utilizing photon energy and converting to thermal energy at interface. Therefore, the lack of fundamental studies related with MXene interfacial energy transport properties would be a hindrance to present the fully potential.

Ultrafast spectroscopy has been employed to study the dynamics of energy pathway for nanomaterials, thus help understanding the mechanisms of electronic and optical properties related applications.[17-19] In the past, many pioneering investigations have provided the crucial dynamics of electronic excitation relaxation in metal nanoparticles[20-23] and some new two-dimensional materials, such as graphene,[24-28] graphene oxide,[29, 30] transition metal dichalcogenides,[31-33] black phosphorus[34-36] or heterostructure 2D materials.[37-39] In principle, with short wavelength pulsed laser generation, the carrier productions of electronic excitation of a nanomaterial will be hot electron-hole pairs.[40] The dissipation processes of the hot carriers are intrinsically converting the absorbed photon energy into other forms of energy in ultrafast time scale. This energy converting pathways could be regulated either by the nanostructures or the different binding molecules.[41] Therefore, the ultrafast studies of energy decay routes of MXene in different interfacial conditions could directly reveal guiding the design of various electronic and thermal related applications through the dissipation pathways.

In this work, we report the comprehensive ultrafast investigation of MXene's dynamic processes of photoexcited carriers in different interfacial conditions, from water to polymers. And the relaxation dynamic processes under plasmonic excitation are separated into chemical bond dominated fast channel and lattice motion dominated thermal channel. The >80% energy of MXene in water could quickly dissipate into attached layer molecules within 3.8 picoseconds (ps) *via* hydrogen bond. The thermal energy taking the remaining part would relax with time constant ~50 ps. The prominent interfacial energy conductance of MXene-water under light stimuli is around 150-300 MW·K$^{-1}$·m$^{-2}$. In ethanol and polymers, energy taken by fast channel decease to 35% with time constant ~9.4 ps. The rest energy in ethanol would dissipate at time constant 400 ps with thermal channel, which would almost 10 times retard the interfacial energy transport. Molecular dynamics results presented that water have much stronger H bonding ability that ethanol on MXene surface. Therefore, our results demonstrated that the effective hydrogen bonding of energy receptors at the surface attached functional groups in the structure of materials plays a key role in the fast energy migration of MXene.

## 2. Results and Discussion

In this study, the MXene nanosheets were prepared by etching MAX phase of Ti$_3$AlC$_2$ using LiF in concert with hydrochloric acid following a previously reported method.[8] After the removal of aluminum atomic layer in Ti$_3$AlC$_2$ by HF, the aluminum sites were replaced by -T groups (-T means -OH, -F, -O-, etc), and the strong Ti-Al metallic bonds were transformed into weak van de waals force, resulting in the easy delamination of MAX phase. As shown in Figure 1a, in Ti$_3$C$_2$T$_x$ nanosheets, titanium atoms are nearly closed packed with carbon atoms occupying octahedral sites to form a five-layer stacked structure with a sequence of Ti/C/Ti/C/Ti, and the end-side titanium atoms were terminated by -T groups. The measured structure in flake shape with around several micrometers is shown in Figure 1b. The main atoms Ti, C, O and some halogen are confirmed by XPS in Figure 1c. In this structure, Ti-C bonds present a mixed covalent/metallic/ionic bonding character, imparting Ti$_3$C$_2$T$_x$ high conductivity (~4600 ± 1100 S/cm) and good plasmonic feature.[42-45] As shown in Figure 1d, MXene aqueous solution exhibits high absorption at 780 nm (1.6 eV), which is attributed to the plasmonic mode.[46]

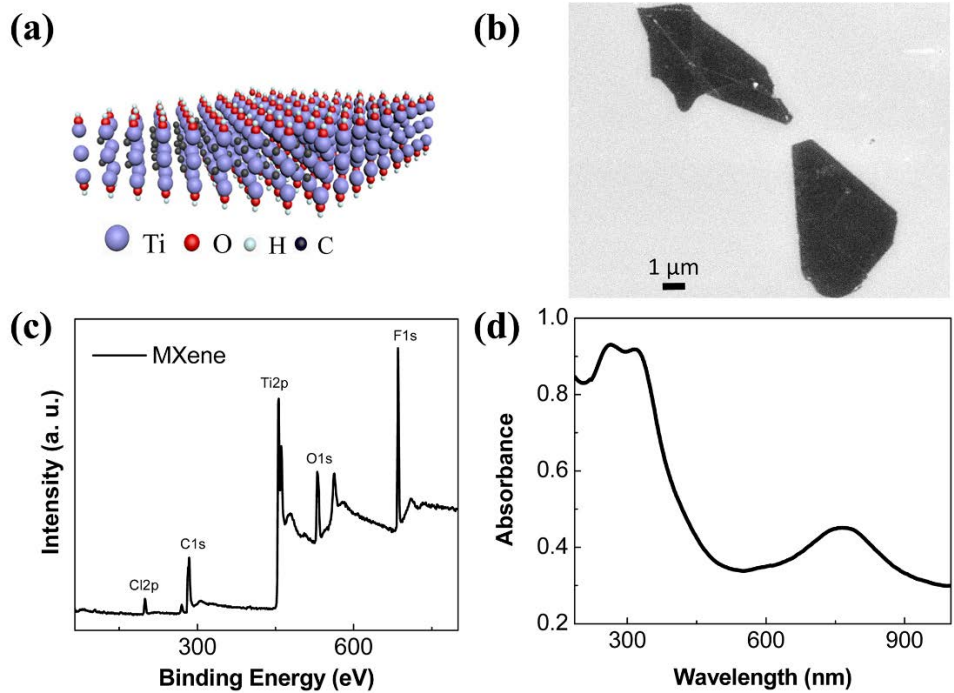

**Figure 1.** Demonstration and characterization of MXene: (a) atomic diagram (b) SEM pictures (c) XPS (d) absorption spectrum of MXene in water solution.

To reveal the detailed carrier and phonon dynamics in MXene after ultrafast plasmonic excitation, the femtosecond pump-probe experiment on MXene nanosheets was first performed. The illumination of MXene with the pump light centered at 800 nm leads to create the transient absorption (TA) spectra at different time delays after the photoexcitation, detecting by the following broadband visible-near infrared light probe pulses. As shown in Fig. 2a, a broadband negative signal with the peak centered at ∼780 nm is clearly observed, which can be attributed to the photobleaching (PB) of the plasmonic absorption band due to the photoexcitation. In addition, a positive absorption band occurs ranging from 450 nm to 700 nm, which is attributed to the photo-induced absorption (PA) signal. The signal intensities of both PB and PA bands remain unchanged after tens of picoseconds time delay. Fig. 2b displays the temporal evolution of the excitation-induced absorption change for the MXene nanosheets detected at different wavelengths, all of which can be well fitted by the triple-exponential function convoluted with an instrument response function (the curves in Fig. 2b). The three fitted time constants are on the femtosecond (∼100 fs), picosecond (1-2 ps) and nanosecond (>10 ns) time scales, respectively. As a consequence, the overall scenario after photoexcitation of MXene nanosheets can be roughly described by three stages with different corresponding time scales, which are marked with different colors in Fig. 2b.

The dynamics that occur following the ultrafast photoexcitation in noble metal nanostructures have been extensively studied by transient absorption experiments. In general, plasmon resonances damping after ultrafast plasmonic excitation would result in the creation of hot electron-hole pairs *via* Landau damping [40]. Then the sequentially occurring dynamics of electron-electron scattering (100 fs), electron-phonon coupling (1-5 ps), and heat dissipation processes *via* coupling to the environment (10-100 ps) are commonly used to describe the ultrafast experimental data. [20-22] In our measurement, the loose contact between the MXene nanosheets and the underlying substrate will inevitably lead to an extremely slow heat dissipation process. Therefore, it is rational to ascribe the three fitted time constants for MXene nanosheets to electron-electron thermalization time (~100 fs), electron-phonon thermalization time (1-2 ps) and heat dissipation processes (>10 ns), respectively.

In comparison, the TA spectra of MXene in aqueous solution are also collected, as displayed in Figures 2c and 2d. A remarkable shorten lifetime is cleared observed, with the intensity of TA spectrum dropped >50% from 0.4 ps to 10 ps. At least four time constants are required to reasonably fit all the dynamic curves in Fig. 2d. The first (~100 fs) and the second (1-2 ps) time constants are consistent with those obtained from MXene nanosheets (Fig. 2b). However, instead of a constant signal at longer time delay shown in Fig. 2b, the dynamics process of MXene in aqueous solution after several picoseconds time delay consists of two decay components: a dominating fast decay process with a time constant about 7 ps (labeled as stage III in Fig. 2d) and a relatively slower decay process with a time constant around 100 ps (labeled as stage III' in Fig. 2d). It indicates that the photo-induced dynamics occurring at longer time delays strongly depends on the surrounding environment of the MXene.

Likewise, the dynamic process of MXene in aqueous solution after plasmonic excitation can be roughly described in three stages, which is illustrated in Fig. 2e. As mentioned above, the first two time constants for MXene in aqueous solution are very close to those for MXene nanosheets. This result is reasonable since the first two stages are dominated by the electron thermalization processes within the MXene lamellae and nearly independent of the surrounding environment, which further confirms our dynamic picture as well as the assignment of corresponding processes. Based on the above analysis, the observed double-exponential decay process in stage III for MXene in aqueous solution represents the rapid heat dissipation dynamics at the interface between the MXene lamellae and the water molecules, during which at least two dissipation channels are involved.

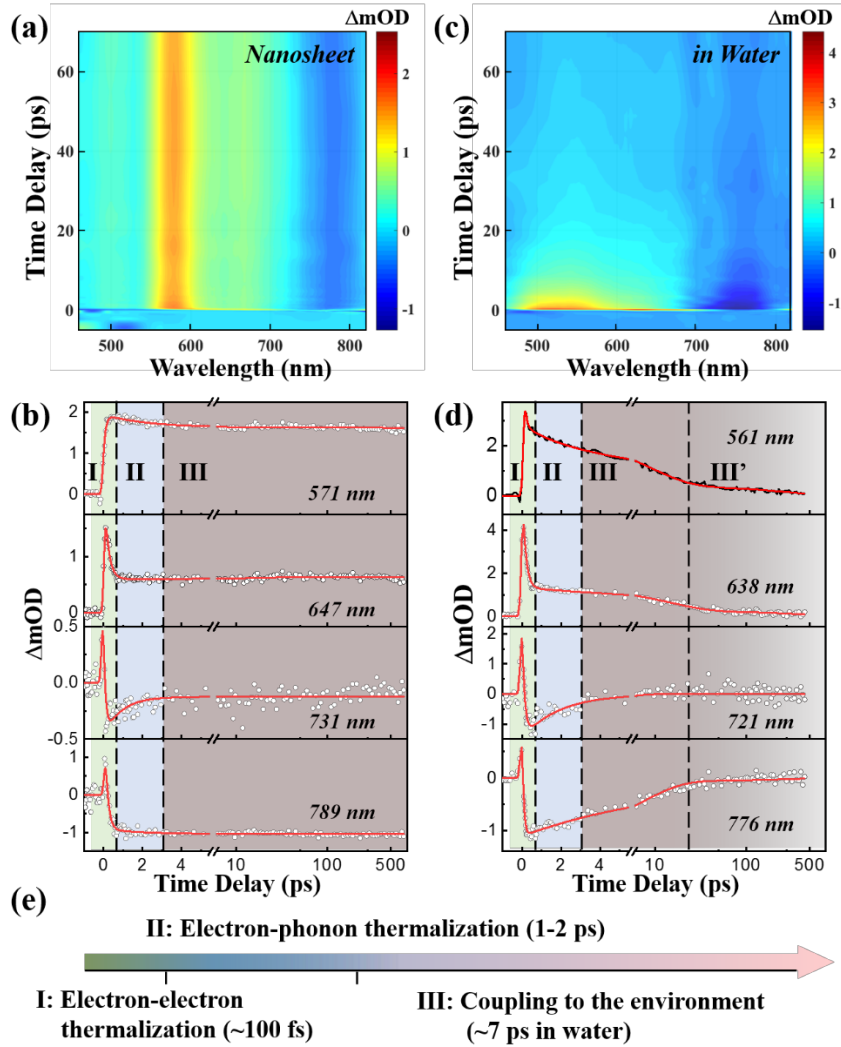

**Fig. 2 | Ultrafast pump-probe measurements of MXene nanosheets and MXene in aqueous solution.** (a) TA spectra of MXene nanosheets measured as a function of time delay after excitation at 800 nm. (b) Temporal evolution of the excitation-induced absorption change for the MXene nanosheets detected at different wavelengths. The dots are the data, and the curves show multiexponential fitting that includes consideration of the instrument response function (~150 fs).

(c) TA spectra of MXene in aqueous solution measured as a function of time delay after excitation at 800 nm. (b) Temporal evolution of the excitation-induced absorption change for the MXene in aqueous solution detected at different wavelengths. The dots are the data, and the curves show multiexponential fitting. (e) Schematic illustration of the overall picture after photoexcitation of MXene. The dynamic process can be roughly described in three stages: (I) electron-electron thermalization, (II) electron-phonon thermalization and (III) coupling to the environment.

To investigate the detailed interfacial heat dissipation channels of MXene in different environments, the TA spectra of MXene samples both dissolved in ethanol solution and mixed in PVA chains were further collected (see Figures 3a and 3b). Similarly, 800 nm excitation pulses also induce the positive PA signal ranging from 450 nm to 700 nm and the negative PB band centered at ~750 nm for both of the two samples. Interestingly, the intensity of the transient spectrum at 25 ps in Fig. 3a for MXene in ethanol still has more than 50% compared with that at 0.5 ps. The temporal evolutions of the normalized excitation-induced absorption change for three MXene samples detected at 520 nm and 780 nm are plotted together in Figures 3c and 3d. Compared with the MXene in aqueous solution, both PA and PB signals for MXene in ethanol solution and PVA mixture exhibit notably increasing lifetimes.

Again, four time constants are employed for fitting all the dynamic curves in Figures 3c and 3d. The first two time constants of ~100 fs and 1-2 ps that contributed by electron-electron and electron-phonon thermalization processes, respectively, are obtained for all three MXene samples in different conditions. In contrast, the relaxation dynamics in stage III strongly dependents on the surrounding environment, and all of them are featured as bi-exponential decays, which we have assigned to two distinct heat dissipation channels after ultrafast thermalization of the MXene. Based on the photoinduced dynamics detected at 520 nm (Fig.3c), the lifetimes for the fast and slow dissipation channels in MXene aqueous solution are fitted as $6.9\pm0.5$ ps and $100\pm10$ ps, respectively, and the amplitude ratio for the fast channel is ~79% (see Figures 3e and 3f). In ethanol solution, the two lifetimes are extended to $33\pm16$ ps and $480\pm80$ ps, respectively, with the amplitude ratio for the fast channel decreased to 24%. In PVA mixture, ~83% energy relaxation flows through the fast channel with time constant 26±1 ps, and the remaining energy dissipates *via* slow channel with time constant 470±2 ps. Comparing the fitting results of MXene in ethanol solution and PVA mixture, it is worth noting that the lifetimes for both the fast and the slow dissipation channels are very similar in two different environments (see Fig. 3e), but the relative ratios of the two decay components are largely different (see Fig. 3f). In all these samples, the ultrafast electronic excitations of MXene perform distinctive relaxation dynamics, indicating different energy transport pathways.

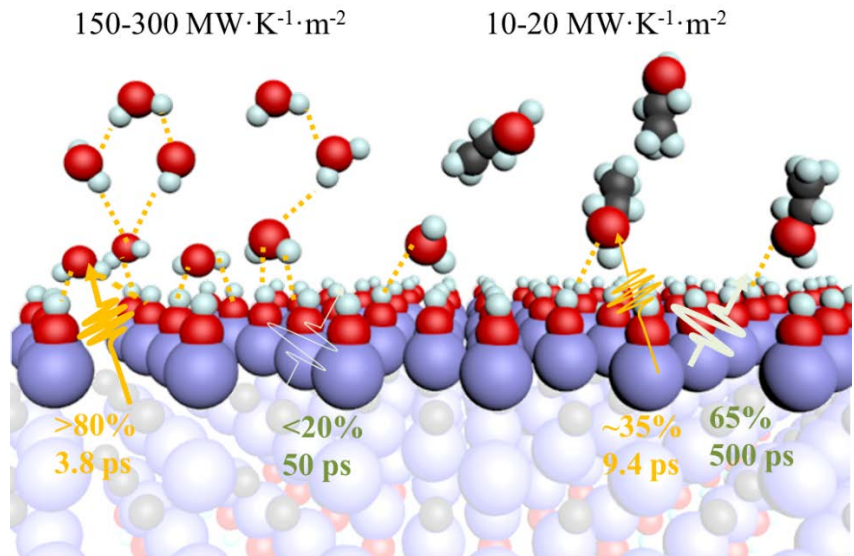

**Figure 4.** Schematic diagram of the entire ultrafast dynamics in solvent samples.

The demonstration of the energy pathways of MXene in solvents are summarized in Figure 4. The studied 2D material $Ti_3C_2T_x$, has large lateral dimensions around several μm (Figure 1b), which could provide multiple solvent-flake interactions. Under resonant illumination, the material absorbs photons and produces the electron-hole pairs. The creation and recombination of electron-hole pairs release energy to dissipate. The surprising results of dynamic behaviors that the interfacial energy dissipation processes occur at $Ti_3C_2T_x$-water interface is significantly faster than that at $Ti_3C_2T_x$-ethanol interface. The dynamics obtained for the two channels thus could be used to roughly calculate the interfacial energy migration. For MXene-water solution with given experimental parameters, after short electron-phonon coupling, all the absorbed energy was converted into internal thermal energy (0.5 μJ/pulse). Then most converted thermal energy *via* hydrogen bond transport out of the MXene flake to interface with time constant 3.8 ps.

With the given surface area of MXene and TiC heat capacity[47], we could then apply Kapitza conductance[48] to calculate the interfacial thermal conductance of MXene-water. Thus, the obtained $G_{MXene-water}$ values varied from 150 MW·K$^{-1}$·m$^{-2}$ to 300 MW·K$^{-1}$·m$^{-2}$.(Details are in the supporting information) For MXene-ethanol solution, 35% energy flux across the interface with time constant 9.4 ps, and 65% energy slowly transfer from MXene to ethanol around 400 ps. Based on this experimental results, the obtained average $G_{MXene-ethanol}$ values were around 10 MW·K$^{-1}$·m$^{-2}$ to 20 MW·K$^{-1}$·m$^{-2}$. (Details are in the supporting information)

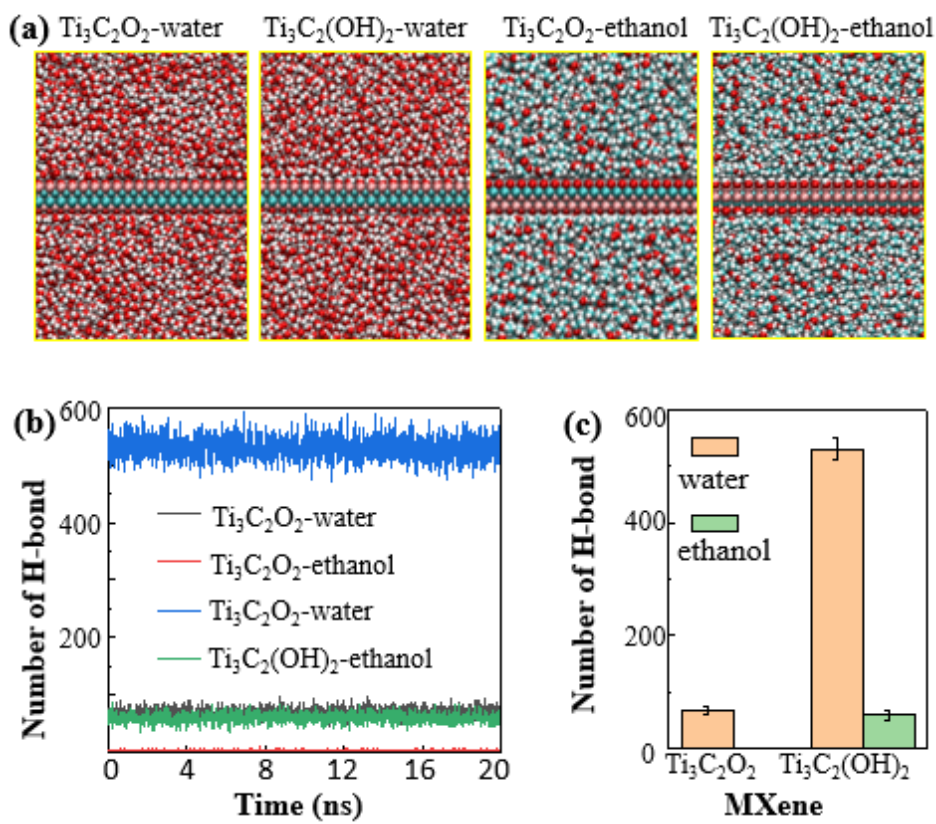

**Figure 5**. MD results for MXene-solvent interface. **(a)** Side-view snapshots (20 ns) for the all-atom systems of $Ti_3C_2O_2$-water, $Ti_3C_2(OH)_2$-water, $Ti_3C_2O_2$-ethanol and $Ti_3C_2(OH)_2$-ethanol. **(b)** Time evolution of the number of h-bonds between $Ti_3C_2O_2/Ti_3C_2(OH)_2$ and water/ethanol. **(c)** Statistical number of H-bonds between $Ti_3C_2O_2/Ti_3C_2(OH)_2$ and water/ethanol over the last 10 ns all-atom MD simulations.

In order to gain the molecular insight, we then employed all atom molecular simulation (MD) for $Ti_3C_2T_x$-solvent interface. Our XPS data showed that the surface group could be either Ti-(OH)$_x$ or Ti-O$_x$ (see Supporting Information figure S2). We selected two typical MXene $Ti_3C_2O_2$ and $Ti_3C_2(OH)_2$ to interact with water and ethanol(figure 5a). We recorded the H-bonding formations on MXene surface with time evolution (figure 5b) from 0-20 ns and counted the

H bonding after 10 ns. As shown in figure 5c, water could form ~9 times more hydrogen bonds than ethanol on Ti$_3$C$_2$(OH)$_2$ surface. The dumping advantage of water also appeared on Ti$_3$C$_2$O$_2$ surface, where H bond ratio of water/ethanol is 69/2. Our results showed that water and ethanol could make significant differences in H bonding formation, indicating the interaction *via* H-bond between MXene surface functional groups and solvent is critical. The pioneers' work had already demonstrated that the number of chemical bond could determine the interfacial thermal conductance.[49] And the stronger interaction with hydrogen bonding could result in larger heat flux across the material-soft interface.[50] In addition, macroscopically, the interaction strength between solvent and MXene could be related to some physicochemical parameters of the solvents, such as dispersion ability, solubility and surface tension. The pioneer's work showed that MXene in water performed better than in ethanol while comparing these parameters, indicating the interaction between water and MXene is stronger.[51] Thus, the combination of our MD and ultrafast dynamics results showed that water have stronger hydrogen bonding ability that ethanol on MXene surface, contributing much more efficient energy transfer pathway for the flake surface. Furthermore, it is interesting to notice that MXene-water thermal conductance value in our work is significantly higher than that other reported typical material-water interfaces, such as octane-water (65 MW·m$^{-2}$·K$^{-1}$),[52] graphene-water (60 MW·m$^{-2}$·K$^{-1}$)[53] and platinum-water (62 MW·m$^{-2}$·K$^{-1}$).[54] As a consequence, our results demonstrated that the 2D structural advantage and hydrophilic groups are critical for the wet MXene to dissipate internal energy out of the plane. Therefore, for effectively converting photon energy into interfacial thermal energy, the crucial way is to construct strong interaction at MXene surface, particularly to find suitable strategy to tightly interact with interfacial groups via hydrogen bonds in wet conditions.

## 3. Conclusion

In summary, the illumination of MXene employing ultrafast laser could generate plasmonic excitation, where the internal decay of electron-hole pair could lead to significant heating of the MXene structure itself. With 2D structural advantage and interfacial interactions, the electronic energy of MXene would more effectively transfer into its environment *via* hydrogen bond. Such fast energy migration could potentially induce the physical and chemical changes of the surrounding environment. Therefore, the present study of the ultrafast dynamics of MXene here has provided a theoretical basis for the control and modification of various types of practical applications, such as photothermal therapeutics, photo acoustic imaging, light induced actuators, harvesting energy conversion from light and so on.

## 4. Materials and methods

### 4.1 Preparation of $Ti_3C_2T_x$ nanosheets (MXene)

The method had been reported by us elsewhere.[55] $Ti_3C_2T_x$ nanosheets were prepared by selectively etching Al atom layer of $Ti_3AlC_2$ MAX ceramic powders (Mesh 200, Micronano Tech. China) using the mixed solution of LiF (Alfa, 98+%) and hydrochloric acid (Greagent, 36-38%). In a typical process, 1g LiF was dissolved into 20 ml of 9 mol/L hydrochloric acid, and then 1g $Ti_3AlC_2$ powders were added into the etchant solution and gently stirred at 35 ºC for 24 h. Next, the products were washed repeatedly with DI water by centrifugation until pH value get close to 6. The centrifugation was performed at the speed of 3500 rpm for 5 minutes each cycle. Then the supernatant was decanted, and the sediment was collected and re-dispersed in DI water by manual shaking for 5 minutes or ultrasonication 2 minute. The dispersion solution was centrifuged again at 3500 rpm for 20 minutes, and the resultant supernatant was decanted and collected for further experiments, tests and characterization. The solid samples were obtained by vacuum drying at 60 ºC or freeze drying for further characterization.

### 4.2 Preparation of stable MXene colloidal solutions with different solvents

In this work, the saturated MXene dispersion solution was prepared by re-dispersing the MXene sediment under ultrasonication (600W, 2 minutes), and then the solution was centrifuged at 3500 rpm for 20 minutes to remove unstable suspensions, and finally stable MXene colloidal solution was obtained. The concentration of stable colloidal solution was measured by weighing the solids in a certain amount of dispersion solution after drying and then calculating their weight percentage, and the MXene solution with specified concentration was prepared by diluting the initial concentrated MXene solution in DI water. For figure 1d, MXene's concentration is around than 0.05mg/mL.

### 4.3 Characterization

UV-vis absorption of MXene solution was measured on a UV-vis spectrometer (UV 2600, Shimadzu, Japan). XPS measurements were performed on an X-ray photoelectron spectrometer (ESCALAB 250Xi, Thermo Scientific) using the Al Kα monochromatic beam (1486.6 eV) with an input power of 150 W. Morphology and microstructures of $Ti_3C_2T_x$ were characterized by a ZEISS Supra55/3195# scanning electron microscope (SEM).

### 4.4 Ultrafast Measurement

Ultrafast spectroscopy measurements were performed by using the output pulses from a femtosecond amplifier laser system (Spitfire Ace, Spectra Physics) with a time duration of ~35 fs, a 800 nm centered wavelength, and a repetition rate of 1 kHz. The laser was split into two beams. The first one was used as the pump beam. 400 nm excitation pulses were further generated by frequency doubled of 800 nm beam with a BBO crystal. The second beam with weaker energy was focused on a sapphire plate to produce broadband white-light continuum pulses as probe pulses. A motorized delay stage was used to control the time delay between the pump and probe beams, both of which were focused onto the sample and overlapped spatially. After passing through the sample, the excitation-induced transmission change of the probe light was collected by an optical fiber spectrometer (AvaSpec-ULS2048CL-EVO, Avantes).

### 4.5 MD Simulation

In this work, CHARMM36m all-atom (AA) force field[56] was used to capture the dynamics of water and ethanol on the surface of MXene ($Ti_3C_2O_2$ and $Ti_3C_2(OH)_2$). The chemical structure and parameters of MXene are adapted or modified from the previous researches.[57, 58] For all simulations, GROMACS software (version 2016.5) and suggested parameters for each force field were used. Each AAMD simulation contained one 9 nm×9 nm MXene sheet, and was run for 20 ns with the time step of 2 fs. The H-bond is counted when the following two conditions are satisfied:(1). The distance between acceptor and hydrogen is less than or equal to distance (default is 3 Å); (2). The angle between donor-hydrogen-acceptor is greater than or equal to angle (default is 120º).

## Acknowledgements

This work is supported by National Natural Science Foundation of China (NSFC-51702009, NSFC-21603270, NSFC-21773302, NSFC-21771017 and NSFC- 21803006) and the Strategic Priority Research Program of Chinese Academy of Sciences (Grant No. XDB30000000). J. L, M. H, G. S and X. L also thank for the Fundamental Research Funds for the Central Universities.

## Additional information

† J. L and Z. C contribute equally;

## Supplementary Information

The Supporting Information is available free of charge from the author.

## Competing financial interests

The authors declare no competing financial interests.